\documentclass[journal]{IEEEtran}
\usepackage{graphicx}
\usepackage[cmex10]{amsmath}
\usepackage{cases}
\usepackage[tight,footnotesize]{subfigure}
\usepackage{amsthm} 
\usepackage{cite}
\usepackage{amssymb}
\usepackage{algorithm}
\usepackage{algorithmic}
\usepackage{multirow}
\usepackage{amsmath}
\usepackage{xcolor}
\usepackage{CJK}
\usepackage{subeqnarray}
\usepackage{cases}
\usepackage{enumerate}
\usepackage{cuted}
\usepackage{booktabs}
\usepackage{mathrsfs}
\usepackage{bbm}
\usepackage{array}

\newcolumntype{M}[1]{>{\centering\arraybackslash}m{#1}}

\usepackage[colorlinks]{hyperref}

\ifCLASSINFOpdf
\else
\fi

\begin{document}

\title{From Fixed to Fluid: Unlocking the New Potential with Fluid RIS (FRIS)}


\author{Han Xiao, Xiaoyan Hu, 
Kai-Kit~Wong,
Xusheng Zhu,  Hanjiang Hong, \\Farshad Rostami Ghadi, Hao Xu, and Chan-Byoung Chae\\
}
\maketitle

\begin{abstract}
Owing to its flexible and intelligent electromagnetic signal manipulation, the technology of reconfigurable intelligent surfaces (RISs) has attracted widespread attention. However, the potential of current RISs can only be partly unlocked due to their fixed geometry and element patterns. Motivated by the concept of the fluid antenna system (FAS), a novel RIS system, termed fluid RIS (FRIS), has been developed. Unlike traditional RISs, FRIS allows the element positions or radiation patterns to exhibit ``fluid" properties, i.e., dynamic reconfigurability, to adapt to the wireless environment, offering enhanced beamforming flexibility and environmental adaptability. Given that research on FRIS is still in its infancy, this paper provides a comprehensive overview of its current developments and future prospects. Specifically, the key features of FRIS are first presented, including its classification, fundamental mechanisms, and advantages. Next, potential application scenarios of FRIS are analyzed and discussed, followed by two illustrative case studies demonstrating its potential. Finally, the main open challenges and future research directions related to FRIS are highlighted.
\end{abstract}
\IEEEpeerreviewmaketitle

\vspace{-4mm}\section{Why FRIS is Necessary for Future Wireless Systems?}\label{sec:S1}
\IEEEPARstart{R}{ecently}, the emerging technology of reconfigurable intelligent surfaces (RIS) has provided a powerful impetus for breakthroughs in beyond-5G (B5G) and 6G wireless networks. By enabling flexible control over the electromagnetic properties of incident signals, RIS transforms wireless systems from passively adapting to the propagation environment into actively reconfiguring it \cite{huang2019reconfigurable}. This paradigm shift has been validated in various wireless applications such as integrated sensing and communications (ISAC) \cite{xu2024intelligent}. However, the current RIS technology only unlocks a fraction of its potentials for enhancing system performance due to the array geometry and element radiation patterns are usually fixed after fabrication in conventional RIS systems, resulting in limited flexibility and  adaptability. 
Additionally, RIS-assisted systems inherently suffer from the multiplicative fading effect. To meet the ever-increasing performance demands, exploring new degrees of freedom (DoFs) of conventional RIS represents a highly promising direction for future applications.

Against this backdrop, the concept of the fluid antenna system (FAS), first proposed by Wong \textit{et al.} in \cite{wong2021fluid, wong2022fluid}, provides a promising avenue for introducing additional DoFs. As a new class of reconfigurable antennas, FAS can dynamically adjust both its position and shape \footnote{From a broad perspective, FAS  also encompasses the frequency-reconfigurable, polarization-reconfigurable, and pattern-reconfigurable antenna systems.}, thereby enriching the physical layer with extra DoFs. And numerous studies have been implemented and validated the potentials of FAS \cite{wee2025channel, hong2025coded}. Inspired by FAS, Ye \textit{et al.} \cite{ye2025joint} introduced position reconfigurability into RIS by enabling the physical movement of its elements, thereby proposing the concept of fluid RIS (FRIS). In addition, simulation results demonstrate that FRIS can significantly enhance system performance by fully exploiting spatial diversity, outperforming traditional RIS.
 Furthermore, \cite{salem2025first, xiao2025fluid} revisited the concept of FRIS by replacing physical movement with virtual movement of elements. This is realized by densely deploying elements across the RIS surface and dynamically activating elements through switching. The potential of this kind of position-reconfigurable FRIS has been validated in both single-user single-input single-output (SU-SISO) and multi-user multiple-input single-output (MU-MISO) scenarios.  A recent study \cite{zhang2025a} has made significant progress in FAS research, showing that within FAS systems, ``reconfigurable position” and ``reconfigurable pattern” can, to some extent, serve functionally interchangeable roles in signal control. Motivated by this insight, we proposed an alternative FRIS paradigm in \cite{xiaofluid_pattern2025}, namely, pattern-reconfigurable FRIS. In this framework, the radiation patterns of the elements will be dynamically adjusted according to the transmission environment, enabling the modification of the electromagnetic properties of incident signals. Additionally, we have shown that FRIS enables path-aware manipulation of incident multi-path signals, fully exploiting the energy carried by the line-of-sight (LoS) and non-line-of-sight (NLoS) components for  constructive signal synthesis, thereby greatly enhancing overall system performance. In contrast, conventional RIS can only perform a uniform adjustment of multi-path signals.

Actually,  research on FRIS is still in its infancy, with numerous challenges yet to be addressed. Moreover, a wide range of promising applications for integrating FRIS into wireless networks remains to be thoroughly explored. Hence, this paper delivers a concise but thorough overview for FRIS and highlights some new insights on its development. In particular, we begin by presenting the classification, signal processing mechanisms, and fundamental advantages of FRIS. We then explore their promising application scenarios in next-generation communication systems, followed by two illustrative case studies showcasing their potentials. Finally, the  key open challenges are highlighted to inspire continued research in this rapidly evolving field.

\section{What is FRIS?}\label{sec:S2}
In this section, an in-depth analysis is provided on the classification, signal processing mechanisms, and core advantages of FRIS
\subsection{Types of FRIS}
\begin{figure*}[!t]
	\centering
	\includegraphics[scale=0.66]{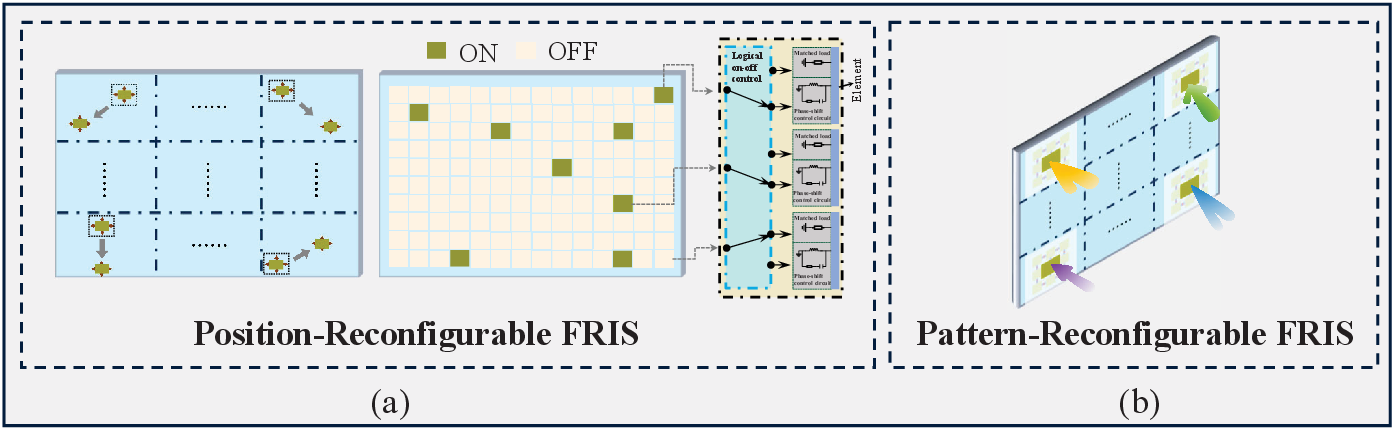}\\
	\caption{Schematic illustrations for different types of FRIS.}\label{fig:Types_of_FRIS}
\end{figure*}

Based on the underlying operating mechanisms, FRIS can be broadly classified into two main categories: \textbf{position-reconfigurable FRIS} \cite{ye2025joint,xiao2025fluid,ghadi2025performance,  ghadi2025fires} and \textbf{pattern-reconfigurable FRIS} \cite{xiaofluid_pattern2025}. These two types differ in their respective reconfiguration strategies for incident electromagnetic signals.  Next, we provide a comprehensive description of these two types of FRIS.
\begin{itemize}
	\item \textbf{Position-Reconfigurable FRIS:} In the position-reconfigurable FRIS framework, which shares the same element structure as the traditional RIS, two types of position-reconfigurable strategies are available, as illustrated in Fig. \ref{fig:Types_of_FRIS} (a). In particular, the first method realizes position reconfigurability through the physical movement of elements on the RIS surface with the help of motors. During this process, the elements can dynamically adjust the arrival phase shifts of signals from the transceivers by leveraging passive beamforming and position reconfigurability, thereby enabling constructive signal combining at the intended receivers. Notably, this reconfigurable strategy is highly constrained in practice, as it may necessitate positional changes on the order of $10$ cm within milliseconds \cite{wang2024ai-enpowered}, corresponding to an acceleration of approximate $2\times 10^5$ m/s$^2$ \footnote{This kind of FRIS with movable elements may be suitable for scenarios where position adjustments are required over a long time scale or can be guided by statistical channel knowledge.}. 
	
	Such extreme dynamics pose significant challenges for practical implementation. In the second scheme, a large number of elements or ports are densely deployed on the RIS surface. Note that each element has two operational modes: `ON' and `OFF'. In the ‘OFF’ mode, the element is connected to a matched load circuit and does not modulate the incident signal. Conversely, in the ‘ON’ mode, the element actively modifies the electromagnetic properties of the incident signals to enhance system performance. By selectively activating a subset of elements at different positions with the assistance of field-programmable gate array (FPGA), the system can actually achieve the virtual movement of elements at an extremely high speed, presenting a significant potential. 	
	
	\item \textbf{Pattern-Reconfigurable FRIS:} Actually, most research efforts in the field of wireless communications have overlooked or underestimated the potential of radiation pattern as a DoF. In this case, our work \cite{xiaofluid_pattern2025} initially introduced the pattern reconfigurability into the RIS systems and proposed a novel FRIS paradigm, i.e., pattern-reconfigurable FRIS, as shown in Fig. \ref{fig:Types_of_FRIS} (b). In this framework, the radiation patterns of the elements dynamically adapt to the transmission environment, enabling the modification of the electromagnetic properties for the incident signals. It is worth pointing out that unlike conventional RIS and position-reconfigurable FRIS, the core of pattern-reconfigurable FRIS lies in the precise control of each element's radiation characteristics. As a result, the element structure must be more sophisticated, often incorporating advanced reconfiguration technologies such as pixel-reconfigurable architectures \footnote{The continuous surface is discretized into small elements, referred to as pixels, with adjacent pixels interconnected via hardwires or RF switches. By dynamically adjusting the connections between neighboring pixels, the pixel-reconfigurable architecture can flexibly modify the electromagnetic characteristics of the elements.} \cite{shen2024antenna} to enable dynamic pattern control.
\end{itemize}
\begin{figure*}[!t]
	\centering
	\includegraphics[scale=0.66]{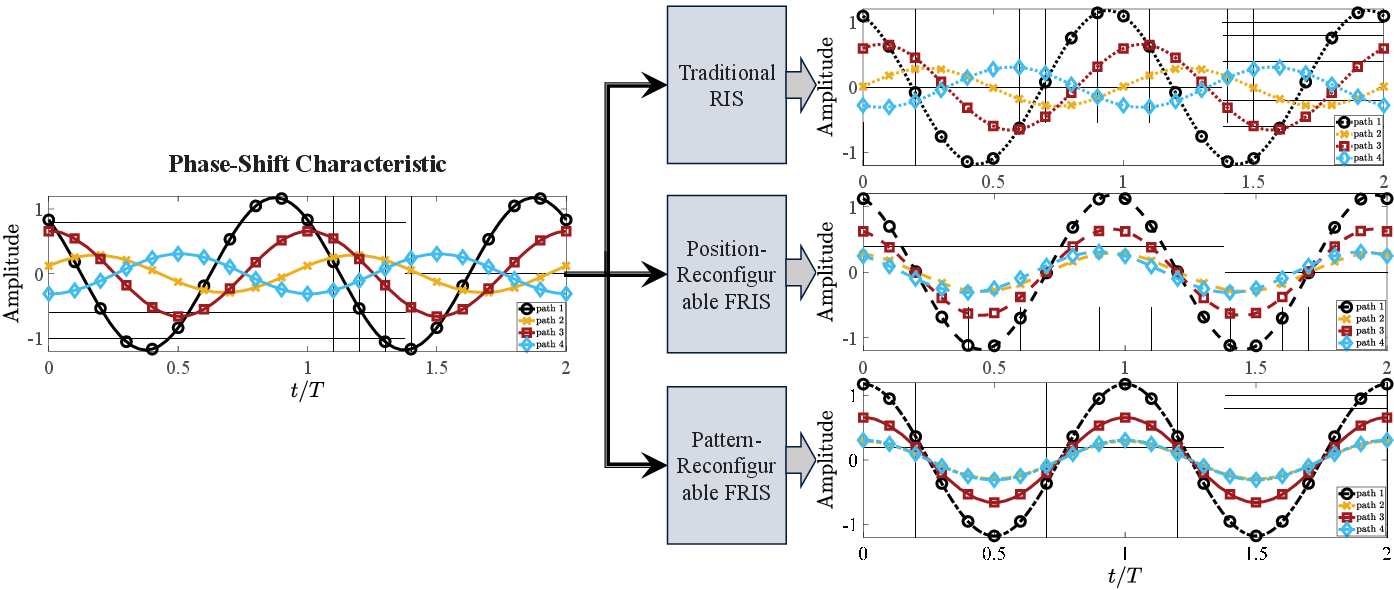}\\
	\caption{Schematic illustrations of signal processing mechanism for different types RIS.}\label{fig:path-aware_schematic}
\end{figure*}
\subsection{Signal Processing Mechanism of FRIS}
This section reveals the signal processing mechanism of FRIS by comparing the received signal power under the traditional RIS and the  FRIS in a multi-path point-to-point down-link communication scenario without LoS links between the BS and the receiver. In the considered scenario, we investigate three types of RIS, i.e., traditional RIS, position-reconfigurable FRIS and pattern-reconfigurable FRIS. The received signal power at the receiver can be expressed as
\begin{align}\label{eq_received_signal_power}
		P=&\frac{1}{LZ}\Bigg|\sum_{m=1}^{M}\vartheta_m\sum_{l=1}^{L}\sum_{z=1}^{Z}g_{l, z}f^{m, l, z}_\mathrm{r, t}(\theta, \phi)\times \notag\\
		&e^{j\frac{2\pi}{\lambda}(\mathbf{k}_{\mathrm{r}, m, l}- \mathbf{k}_{\mathrm{t}, m, z})^T\mathbf{p}_m}\Bigg|^2+\sigma^2,
\end{align}
where $L$ and $Z$ denote the number of multi-path components from the BS to the RIS and from the RIS to the receiver, respectively. $M$ is the number elements equipped at the RIS. $\vartheta_m, ~m\in\mathcal{M}\triangleq\{1, \cdots, M\}$ represents the reflection coefficient of the $m$-th element. $g_{l, z}$ denotes the complex channel gain of the $(l, z)$-th path of the cascaded channel from the BS to the receiver. $f^{m, l, z}_\mathrm{r, t}(\theta, \phi)$ is the radiation pattern gain of the $m$-th element in the direction corresponding to the $(l, z)$-th path of the cascaded channel. $\mathbf{p}_m$ represents the position vector of the $m$-th element. $\mathbf{k}_{\mathrm{r}, m, l}$ and $\mathbf{k}_{\mathrm{t}, m, z}$ are the normalized wave vectors. $\sigma^2$ is the noise power. Note that in pattern-reconfigurable FRIS systems, the reflection coefficients are typically fixed or set to one. In both traditional RIS and pattern-reconfigurable FRIS-aided systems, the element positions remain static. Conversely, in traditional RIS and position-reconfigurable FRIS-assisted systems, the radiation patterns of the elements are fixed or assumed to be isotropic.

To maximize the received signal power in \eqref{eq_received_signal_power}, we can observe that the reflection coefficients in traditional RIS systems can only be adjusted to align with the overall phase shift of the cascaded channel. In position-reconfigurable FRIS, the coupling between element positions and normalized wave vectors enables individual adjustment of the phase shifts for each multi-path component. However, the modulation capability becomes limited, especially as the number of multi-path components increases, since it is challenging to find position configurations that satisfy the optimal conditions within a confined space without overlap. Passive beamforming can be employed to complement this limitation and improve overall performance.
Since the radiation pattern is a function of 3D spatial parameters, it can theoretically provide enough DoFs to effectively modulate the phase shifts of individual multi-path components at different directions, thereby enabling constructive signal combining at the receiver. Moreover, since the shape of each element’s radiation pattern determines how signal energy is distributed across spatial directions, this shape-based DoF can be leveraged to significantly boost the modulation flexibility in pattern-reconfigurable FRIS systems. 

A simple example is utilized to validate the path-aware modulation capability of the FRIS. In particular, we first randomly generate four multi-path components considering parameters $L=1, Z=4$. The left side of Fig.~\ref{fig:path-aware_schematic} illustrates the phase-shift features of these four multi-path components. Note that, the primary objective is to verify the phase-shift modulation capability of the FRIS, the large-scale path loss of all multi-path components is temporarily neglected. Then, the traditional RIS and FRIS are used to manipulate these four multi-path components with the aim of maximizing the received signal power, respectively. The right panel of Fig.~\ref{fig:path-aware_schematic} depicts the phase-shift characteristics of the four multi-path components after modulation with three types of RIS, enabling a direct comparison between the traditional RIS and the FRIS. According to the presented results, it is observed that under the control of a traditional RIS, the relative positions of the four multi-path components in the figure remain unchanged, undergoing only a collective phase translation. In contrast, with the position-reconfigurable FRIS, the four multi-path components are adjusted as much as possible toward in-phase alignment, demonstrating its path-aware modulation capability. Furthermore, under the pattern-reconfigurable FRIS, the four multi-path components achieve nearly perfect in-phase alignment, clearly evidencing its stronger path-aware modulation capability and substantial potentials.
Reference \cite{xiaofluid_pattern2025} presents a more detailed discussions on the principles and potential of FRIS.
\begin{table*}[h!]
    \renewcommand\arraystretch{1.4}
    \centering
    \caption{Comparison of FRIS and Conventional RIS}
    \label{tab:table1}
    \begin{tabular}{p{2.8cm} p{2.8cm} p{3.2cm} p{3.2cm} p{3.2cm}}
        \toprule
        \textbf{Types} & 
        \textbf{Optimization DoFs} & 
        \raggedright\textbf{Manipulation of Multi-path Signals} & 
        \raggedright\textbf{Performance Enhancement Potential} & 
        \textbf{Implementation Complexity} \\
        \midrule
        Traditional RIS & Passive beamforming & Overall manipulation & Low & Low \\
        \midrule
        Position-reconfigurable FRIS & Passive beamforming + Position & Quasi-path-aware & High & First type: High; Second type: Medium \\
        \midrule
        Pattern-reconfigurable FRIS & Radiation pattern & Path-aware & Very high & High \\
        \bottomrule
    \end{tabular}
\end{table*}

\subsection{Key Advantages of Employing FRIS}
Based on the above analysis and discussion of the FRIS mechanism, integrating FRIS into wireless communication systems offers the following advantages over conventional RIS.
\begin{itemize}
	\item \textbf{Multi-path Diversity Gain:} Considering that the conventional RIS can only perform overall adjustment of the cascaded channel, its reflection coefficients are typically designed to align with the virtual LoS path between the transceivers so as to maximize system communication performance, as the LoS path carries the largest portion of the signal energy. In other words, the traditional RIS tends to exploit only the strongest path, while weaker multi-path components are often ignored, wasted, or even suppressed. This limitation leads to inefficient energy utilization. In contrast, FRIS offers path-aware control over multi-path channels, allowing it not only to fully exploit the energy carried by the LoS path but also to incorporate NLoS components as auxiliary energy sources for constructive signal synthesis, thereby significantly enhancing the overall system performance gain.
	\item \textbf{Flexible Beamforming:} For conventional RIS systems, once fabricated, both the array geometry and the element radiation patterns are fixed. This inherent rigidity limits the flexibility and adaptability of beamforming in varying communication environments, often preventing the system from achieving optimal beam control under diverse wireless channel conditions.	
	In contrast, the position-reconfigurable FRIS can dynamically alter its array geometry by adjusting the physical positions of its elements, thereby achieving spatial reconfigurability at the physical layer. Combined with its inherent passive beamforming capability, the position-reconfigurable FRIS can more flexibly control the phase distribution of signals, enabling customized beam optimization tailored to different user locations and multi-path structures.	
	Although the geometry of the pattern-reconfigurable FRIS remains unchanged, its advantage lies in the tunability of the element radiation patterns. In addition to phase-shift control, variations in the pattern shape can directly influence the energy distribution of different multi-path components, allowing differentiated weighting or suppression of their contributions. This dual control over both phase and energy distribution endows the pattern-reconfigurable FRIS with notable beamforming flexibility and environmental adaptability.
Therefore, both types of FRIS break through the inherent beamforming limitations of conventional RIS from different dimensions.
\end{itemize}

Table \ref{tab:table1} compares conventional RIS and the two types of FRIS in terms of optimization DoFs, ability to manipulate multi-path signals, performance improvement potential, and implementation complexity.

\begin{figure*}[!t]
	\centering
	\includegraphics[scale=0.85]{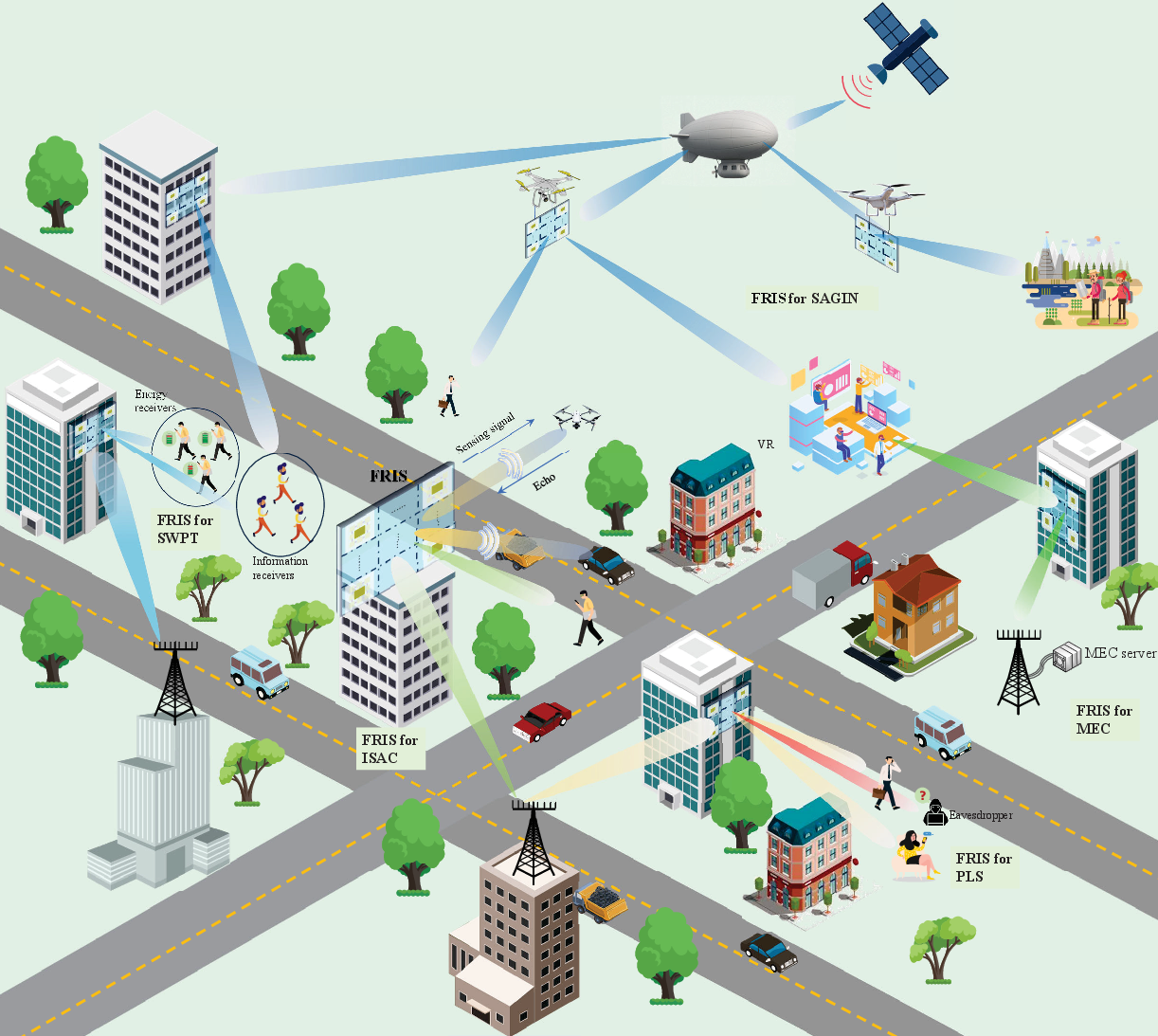}\\
	\caption{A forward-looking vision of FRIS seamlessly integrated into wireless networks for a broad range of applications.}\label{fig:Scenario}
\end{figure*}

\section{Promising Applications of FRIS for 6G and Beyond}\label{sec:S3}
Actually, with its path-aware manipulation capability, FRIS offers significant opportunities to boost system performance In this section, we discuss its promising applications in next-generation communication systems.
\subsection{FRIS for ISAC}
As one of the key enabling technologies for future 6G networks, ISAC technology enables simultaneous environment sensing and data transmission using the same spectral resources, thereby improving the efficiency of spectrum, hardware, energy, and cost. However, implementing ISAC faces a fundamental challenge: communication and sensing require distinct beamforming patterns. Communication typically demands a highly directional beam aimed at the user, whereas sensing benefits from a wide beam or multiple directional beams to detect and track targets. This disparity imposes stricter requirements on the beamforming capability of systems. Compared with conventional RIS, FRIS enable position reconfiguration or radiation pattern reconfiguration, which effectively expands the spatial DoFs and enhances system flexibility. This capability allows the system to dynamically tailor its beam patterns—forming narrow beams to provide high-gain, focused communication links while simultaneously maintaining wide-angle or multi-directional beams to support reliable target sensing. 

Furthermore, we further analyze how FRIS mitigate this challenge from a channel-energy perspective. In practical deployments, particularly in non-cooperative or passive sensing scenarios, the NLoS components between the RIS and the target are often difficult to acquire directly. As a result, target sensing primarily relies on the LoS path. Although the signal energy carried by NLoS components cannot be exploited for sensing under such conditions, the flexible reconfigurability of FRIS introduces additional DoFs, enabling precise control over the signal energy distribution along the LoS path and thereby significantly enhancing sensing performance.
For communications, the path-aware manipulation capability  of FRIS allows full utilization of both LoS and NLoS signal components reaching the user, leading to substantial improvements in communication capacity.
Therefore, FRIS empowers ISAC systems to simultaneously achieve breakthroughs in both sensing accuracy and communication performance.
\subsection{FRIS for PLS}
Due to the open nature of the wireless transmission environment, wireless channels are highly susceptible to security threats such as jamming attacks and information leakage. To address these security challenges, physical layer security (PLS), grounded in information theory, has emerged as a promising solution, as it can avoid complex key exchange protocols and is well-suited for latency-sensitive applications. However, the majority of existing PLS transmission schemes, including conventional RIS-aided PLS schemes, offer only limited upper bound on secure performance.
Owing to FRIS's path-aware manipulation capability, it can further unlock the potential of PLS techniques. Specifically,  multi-path signals can be coherently combined at the legitimate users to improve the signal-to-interference-plus-noise ratio (SINR), while destructive combining of multi-path components can be induced at the eavesdroppers to substantially reduce the received signal energy. By this means, information leakage risks can be effectively mitigated, even when eavesdroppers are located in close proximity to the FRIS.
\subsection{FRIS for SWIPT}
Simultaneous wireless information and power transfer (SWIPT) is a promising technique for future energy-constrained Internet of Things (IoT) networks, in which part of the transmission signal energy is used to charge IoT devices, while the remaining energy is utilized for information transmission. In SWIPT systems, a critical issue must be considered: to ensure the stable operation of devices, the required energy may be much higher than that needed for information transmission. Due to severe path loss, the signal energy arriving at the devices is highly limited, leading to a significant trade-off for the allocation of received signal energy. Moreover, the limited signal energy also imposes strict constraints on the achievable coverage area. FRIS can effectively address this limitation with the assistance of its path-aware modulation ability. In particular, under FRIS manipulation, the LoS and NLoS components can be constructively combined at the devices, enabling the signal energy carried by both components to be efficiently delivered to the devices and thereby mitigating the challenge faced by current SWIPT schemes, which is unachievable with conventional RIS. 
\subsection{FRIS for MEC}
With the continuous advancement of communication technologies, emerging applications such as virtual reality (VR) and pervasive gaming require wireless networks to handle computation-intensive and latency-critical tasks in real time, given the limited power supply and hardware capabilities of local devices. This challenge can be effectively addressed through MEC, which relocates computational capacity from the central cloud to the edge of the network. However, in some special scenarios where terminal devices are located far from MEC nodes or lack LoS links to them, the offloading capability is severely constrained. FRIS provides a promising solution for such scenarios, as it not only establishes virtual LoS links for MEC networks but also enables receivers to fully exploit the signal energy carried by multi-path components, thereby significantly improving SINR and enhancing offloading capacity.
\subsection{FRIS for SAGIN}
Space-air-ground integrated network (SAGIN), which interconnects space, air, and ground segments, enables seamless communication across diverse environments and supports applications such as global connectivity, broadband Internet access, remote sensing, disaster management, and military operations. However, SAGIN still suffers from some challenges such as signal coverage and capacity in harsh environments, propagation losses in the atmosphere and space, and high energy consumption. Fortunately, FRIS can effectively overcome these limitations by utilizing its flexible beamforming ability and path-aware signal manipulation. Specifically, FRIS can dynamically reconfigure the electromagnetic characteristics of signals to extend the coverage of wireless transmissions, while coherently combining the multi-path components at desired locations to enhance signal strength.


\section{Case Study}\label{sec:S4}
This section examines two cases, position-reconfigurable FRIS and pattern-reconfigurable FRIS assisted wireless networks, to demonstrate the potential of FRIS in enhancing system performance through comparison with baseline schemes.
\subsection{Position-Reconfigurable FRIS}
In first case, we explore the potential of the second type of position-reconfigurable FRIS by considering a classical SU-SISO system (see \cite[Fig. 1]{xiao2025fluid}) where each element equipped at the FRIS can only provide a limited phase-shift resolution for the modulation of incident signals. To evaluate the performance gain of FRIS, an optimization problem with the aim of maximizing the achievable rate taking account of position selection of activated elements in FRIS and their discrete phase-shift constraints is formulated. To address this discrete problem, an effective algorithm based on the cross-entropy optimization (CEO) framework is proposed to design all the variables without resorting to alternative strategies. Note that the benchmark scheme associated with traditional RIS is utilized in which the elements are uniformly distributed across the available region.
\begin{figure}[t]
	\centering
	\includegraphics[scale=0.55]{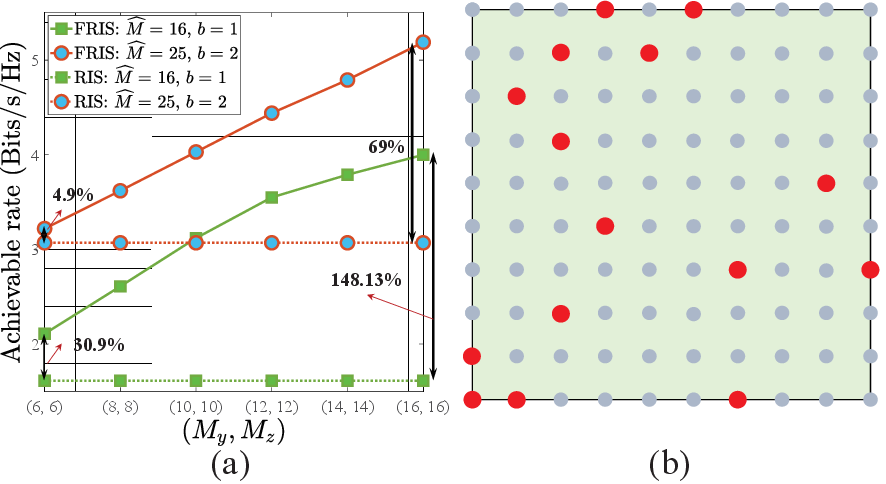}\\
	\caption{ The obtained simulation results: (a) The achievable rate versus the number of elements equipped at the	FRIS system considering the different number of activated elements and phase-shift resolution; (b) the optimal position-reconfigurable FRIS configuration. The detailed simulation setups are presented in \cite{xiao2025fluid}.}\label{fig:case1}
\end{figure}

Fig. \ref{fig:case1} (a) investigates the influence of the number of elements equipped at the FRIS system on user's achievable rate considering different number of activated elements $\widehat{M}$ and phase-shift resolution $b$. It is worth pointing out that $\widehat{M}$ denotes the total number of elements in traditional RIS-aided scheme.
From this figure, we can find that the achievable rate grows with the number of elements equipped on the FRIS across all scenarios, since the positions of activated elements have the wider selection to offer greater spatial DoFs for signal modulation. Moreover, simulations indicate that, under identical conditions, the proposed FRIS framework substantially outperforms conventional RIS, and this performance advantage becomes increasingly evident as the number of available positions expands. In particular, when $\widehat{M}=16$ and $b=1$, the gain increases from $30.9\%$ to $148.13\%$ as the number of elements grows from $6 \times 6$ to $16 \times 16$. Similarly, when $\widehat{M}=25$ and $b=2$, the gain rises from $4.9\%$ to $69\%$ over the same element range, demonstrating the strong potential of FRIS to significantly enhance wireless network performance. Furthermore, Fig. \ref{fig:case1}(b) illustrates the optimal position selection of activated elements, considering the case where $16$ elements are activated in a $10 \times 10$ configuration.
\subsection{Pattern-Reconfigurable FRIS}
This section focuses on exploring the potential of pattern-reconfigurable FRIS by integrating it into a multi-user communication system (\cite[Fig. 1]{xiaofluid_pattern2025}). The performance gain introduced by the pattern-reconfigurable FRIS is evaluated using the weighted sum rate. Furthermore, we formulate an optimization problem aimed at maximizing the weighted sum rate of the users by taking into account the power budget at the base station (BS), and the energy constraints of elements' radiation patterns. To quantify the performance gain achievable by the pattern-reconfigurable FRIS, the following baseline schemes associated with the traditional RIS are employed: \textbf{1) Traditional RIS with 38.901 pattern scheme:} In this scheme, the patterns of elements are fixed as 3GPP 38.901 pattern; \textbf{2) Traditional RIS with isotropic pattern scheme}.
\begin{figure}[t]
	\centering
	\includegraphics[scale=0.85]{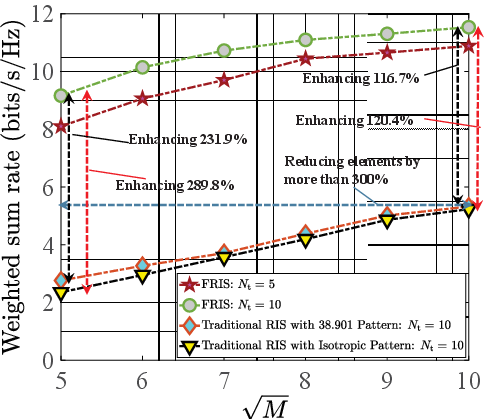}\\
	\caption{ The weighted sum rate versus the number of elements equipped at the FRIS system considering the different number of antennas. The detailed simulation setups are presented in \cite{xiaofluid_pattern2025}.}\label{fig:case2}
\end{figure}

Fig.~\ref{fig:case2} illustrates how the weighted sum rate varies with the number of elements, $M$, deployed at the FRIS, under different antenna configurations. Specifically, to ensure a clear performance comparison, the case of BS with the number of antennas $N_\mathrm{t} = 10$ is utilized to implement the traditional RIS-based schemes considering 38.901 pattern and isotropic pattern. Simulation results show that even compared to the FRIS scheme with $N_\mathrm{t} = 5$, the traditional RIS scheme using the 38.901 and isotropic radiation patterns perform significantly worse. Under the same conditions, the proposed scheme achieves improved performance gains ranging from $116.7\%$ to $231.9\%$ and $120.4\%$ to $289.8\%$ over traditional RIS with the 38.901 and isotropic patterns, respectively, as the number of elements varies from $100$ to $25$. The average performance improvements are $161.5\%$ and $176.2\%$, respectively, highlighting the strong potential of the pattern-reconfigurable FRIS in enhancing system performance. Moreover, compared to the traditional RIS scheme, the FRIS-assisted schemes can reduce the required number of elements by more than $300\%$ while maintaining the same performance (e.g., FRIS with $M = 25$ elements versus traditional RIS with $M = 100$ elements). This demonstrates the remarkable hardware efficiency of FRIS.


\section{Challenges and Research Directions}\label{sec:S5}
Given that FRIS research is still in its early stages, numerous open challenges must be addressed to fully realize its potential. This section highlights several key challenges and prospective research directions for the practical implementation of FRIS.
\subsection{Channel Estimation}
 The key to unlocking FRIS’s full potential lies in the rapid and accurate acquisition of channel state information (CSI). Specifically, unlike conventional RIS, which typically requires the estimation of a single cascaded channel matrix, FRIS must estimate multiple distinct channel states to account for varying position configurations and radiation patterns, resulting in substantial pilot overhead and time cost. Thus, developing effective and low-complexity channel estimation algorithms are the promising research direction for the practical implementation of FRIS.
\subsection{Hardware Design}
Compared with traditional RIS, FRIS offers additional DoFs for manipulating incident signals, but this advantage comes at the cost of increased hardware implementation complexity. During the process of implementing the first type of position-reconfigurable FRIS, FRIS systems face challenges such as high-precision actuation and positioning feedback, mechanical durability, increased power consumption, and coordinated control of multiple. The second type of position-reconfigurable FRIS, which relies on position selection rather than frequent mechanical movement, still encounters several hardware challenges. These include designing switch matrix circuit with low insertion loss, low power consumption, and high isolation; addressing routing complexity and parasitic coupling effects; enabling efficient control of large-scale switch arrays; and enhancing adaptability to dynamic environmental conditions. Unlike conventional RIS and position-reconfigurable FRIS, the primary hardware challenge of pattern-reconfigurable FRIS lies in developing elements capable of dynamically reconfiguring their radiation patterns. At present, very few studies address the design of such advanced elements, leaving this area largely unexplored. As a result, advancing the hardware implementation of pattern-reconfigurable FRIS stands out as a highly promising and valuable research direction. Overall, overcoming these hardware challenges requires collaborative multidisciplinary efforts.
\subsection{AI-Driven Optimization Design}
Although the powerful signal manipulation capabilities of FRIS can substantially enhance communication system performance, they also introduce additional design variables, i.e., position optimization and pattern optimization, that are not present in conventional RIS. Additionally, in practical deployments, FRIS is often required to equip with hundreds of elements to effectively control incident signals. Therefore, model-based optimization algorithms, which typically require numerous iterations to obtain near-optimal solutions due to the non-convexity of their objective functions and constraints, face extremely high computational complexity, posing a significant barrier to real-time applications. The artificial intelligence (AI)-based approach offers a promising solution to the aforementioned challenge. Being the data-driven method, it can extract system features without relying on explicit mathematical models, and once trained, it can generate near-optimal solutions through simple algebraic computations, even in the presence of imperfect CSI and hardware impairments.

\section{Conclusion}\label{sec:S6}
This article reviewed a novel RIS technology, fluid RIS (FRIS), which includes both position-reconfigurable and pattern-reconfigurable types. First, the basic principles and key advantages of FRIS over conventional RIS have been analyzed. Building on these advantages, the promising applications of FRIS in wireless communication networks have been explored. Furthermore, two numerical simulation cases, one for position-reconfigurable FRIS and one for pattern-reconfigurable FRIS, have been presented to demonstrate the potential advantages of FRIS compared to traditional RIS. Finally, the main open challenges and future research directions related to FRIS have been discussed.

\ifCLASSOPTIONcaptionsoff 
  \newpage
\fi
\bibliographystyle{IEEEtran}

\section*{Biographies}
\textsc{Han Xiao} (Graduate Student Member, IEEE) is a PhD candidate at the School of Information and Communications Engineering, Xi'an Jiaotong University.
\vspace{2mm}

\textsc{Xiaoyan Hu} (Member, IEEE) is a Professor at the School of Information and Communications Engineering, Xi'an Jiaotong University.
\vspace{2mm}

\textsc{Kai-Kit Wong} (Fellow, IEEE) is a Professor at the Department of Electronic and Electrical Engineering, University College London and also a Professor at Yonsei Frontier Lab, Yonsei University, Seoul.
\vspace{2mm}

\textsc{Xusheng Zhu} (Member, IEEE) is a  Post-Doctoral Researcher at the Department of Electronic and Electrical Engineering, University College London.
\vspace{2mm}

\textsc{Hanjiang Hong} (Member, IEEE) is a  Post-Doctoral Researcher at the Department of Electronic and Electrical Engineering, University College London.
\vspace{2mm}

\textsc{Farshad Rostami Ghadi} (Member, IEEE) is a  Post-Doctoral Researcher at the Department of Electronic and Electrical Engineering, University College London.
\vspace{2mm}

\textsc{Hao Xu} (Senior Member, IEEE) is a Professor at National Mobile Communications Research Laboratory, Southeast University, Nanjing.
\vspace{2mm}

\textsc{Chan-Byoung Chae} (Fellow, IEEE) is a Professor at School of Integrated Technology, Yonsei University, Seoul.
\vspace{2mm}
\end{document}